\documentclass{article}

\usepackage{arxiv}

\usepackage[utf8]{inputenc} 
\usepackage[T1]{fontenc}    
\usepackage{hyperref}       
\usepackage{url}            
\usepackage{booktabs}       
\usepackage{amsfonts}       
\usepackage{nicefrac}       
\usepackage{microtype}      
\usepackage{lipsum}		
\usepackage{graphicx}
\usepackage{doi}
\usepackage{acronym}

\usepackage{siunitx}
\usepackage[numbers]{natbib}
\usepackage{tikz}       
\usetikzlibrary{shapes}
\usepackage{xcolor}     

\usepackage[abs]{overpic}
\usepackage{hyperref}
\usepackage{multirow}%
\usepackage{amsmath,amssymb,amsfonts}%
\usepackage{amsthm}%
\usepackage{bm}
\usepackage{mathrsfs}%
\usepackage[title]{appendix}%
\usepackage{xcolor}%
\usepackage{textcomp}%
\usepackage{manyfoot}%
\usepackage{booktabs}%
\usepackage{algorithm}%
\usepackage{algorithmicx}%
\usepackage{algpseudocode}%
\usepackage{acronym}
\usepackage{listings}%
\usepackage{scalerel}
\usetikzlibrary{svg.path}
\usepackage{algorithm}
\usepackage{algpseudocode}

\usepackage{graphicx}    
\usepackage{overpic}     

\acrodef{EPOD}[EPOD]{Extended Proper Orthogonal Decomposition}
\acrodef{GM}[GM]{Galerkin model}
\acrodef{PIV}[PIV]{Particle Image Velocimetry}
\acrodef{POD}[POD]{Proper Orthogonal Decomposition}
\acrodef{DNS}[DNS]{Direct Numerical Simulations}
\acrodef{CFD}[CFD]{Computational Fluid Dynamics}
\acrodef{LES}[LES]{Large Eddy Simulation}
\acrodef{LOR}[LOR]{low-order reconstruction}
\acrodef{PDE}[PDE]{partial differential equation}
\acrodef{ODE}[ODE]{ordinary differential equation}
\acrodef{RANS}[RANS]{Reynolds-Averaged Navier-Stokes}
\acrodef{ROM}[ROM]{Reduced-Order Model}
\acrodefplural{ROM}[ROMs]{Reduced-Order Models}
\acrodef{SVD}[SVD]{Singular Value Decomposition}
\acrodef{TR}[TR]{time-resolved}
\acrodef{TR PIV}[TR PIV]{time-resolved PIV}
\acrodef{TSS}[TSS]{time super-sampling}
\acrodef{NTR}[NTR]{non-time-resolved}
\acrodef{PSD}[PSD]{Power spectral density}

\title{Model-based time super-sampling of turbulent flow field sequences}

\author{ \href{https://orcid.org/0009-0008-5601-9970}{\includegraphics[scale=0.06]{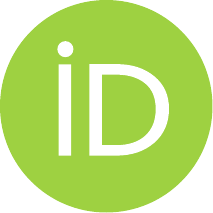}\hspace{1mm}Qihong L. Li-Hu}\\
    Department of Aerospace Engineering\\
    Universidad Carlos III de Madrid\\
     Leganés, Madrid, Spain\\
	\texttt{qihonglorena.li@uc3m.es} \\
	\And
	\href{https://orcid.org/0000-0002-0735-4579}{\includegraphics[scale=0.06]{orcid.pdf}\hspace{1mm}Patricia García-Caspueñas} \\
	Department of Aerospace Engineering\\
    Universidad Carlos III de Madrid\\
     Leganés, Madrid, Spain\\
    \And
	\href{https://orcid.org/0000-0001-7342-4814}{\includegraphics[scale=0.06]{orcid.pdf}\hspace{1mm}Andrea Ianiro} \\
	Department of Aerospace Engineering\\
    Universidad Carlos III de Madrid\\
     Leganés, Madrid, Spain\\
    \And
	\href{https://orcid.org/0000-0001-9025-1505}{\includegraphics[scale=0.06]{orcid.pdf}\hspace{1mm}Stefano Discetti} \\
	Department of Aerospace Engineering\\
    Universidad Carlos III de Madrid\\
     Leganés, Madrid, Spain\\
}

\renewcommand{\shorttitle}{Model-based time super-sampling of turbulent flow fields sequences}

\hypersetup{
pdftitle={Model-based time super-sampling of turbulent flow field sequences},
pdfsubject={physics.flu-dyn},
pdfauthor={Qihong L. Li-Hu, Patricia García-Caspueñas, Andrea Ianiro, Stefano Discetti},
pdfkeywords={Time super-sampling, POD, Galerkin model, ROM},
}

\begin{document}
\maketitle
\acresetall

\begin{abstract}
We propose a novel method for model-based time super-sampling of turbulent flow fields. The key enabler is the identification of an empirical Galerkin model from the projection of the Navier-Stokes equations on a data-tailored basis. The basis is obtained from a \ac{POD} of the measured fields. Time super-sampling is thus achieved by a time-marching integration of the identified dynamical system, taking the original snapshots as initial conditions. Temporal continuity of the reconstructed velocity fields is achieved through a forward-backwards integration between consecutive measured Particle Image Velocimetry measurements of a turbulent jet flow. The results are compared with the interpolation of the \ac{POD} temporal coefficients and the low-order reconstruction of data measured at a higher sampling rate. In both cases, the results obtained show the ability of the method to reconstruct the dynamics of the flow with small errors during several flow characteristic times.
\end{abstract}

\acresetall

\section{\label{sec:introduction}Introduction}
Knowledge of fluid flow dynamics remains today as the paramount constituent of fluid system modelling, weather forecasting, flow control and design optimization, among others. Access to flow fields with sufficient temporal resolution enables a detailed understanding of its flow dynamics \cite{brunton2015closed}. Technological limitations bound this possibility to a narrow range of flow conditions. In this work, we propose a \ac{TSS} method to model the dynamics and enhance the temporal resolution of experimental and numerical data.

\ac{CFD} \citep{anderson1995computational} comprises a numerical retrieval approach for turbulent flow dynamics. In this way, the Navier-Stokes equations can be discretised to retrieve \ac{TR} flow fields, with methods ranging from \ac{LES} to \ac{DNS}. Nonetheless, \ac{CFD} simulations are computationally expensive \citep{pope2001turbulent} and become inaccessible for moderate to high Reynolds numbers ($Re$). On the other hand, experimental measurements are more suitable to tackle flow dynamics description at higher $Re$, either qualitatively with flow visualisation \citep{merzkirch2012flow} or quantitatively, with \ac{TR} flow field measurements, such as \ac{TR PIV} \citep{beresh2021time}.

In spite of its fast-paced advancements, the application of \ac{TR PIV} is still limited to flows at low-to-moderate velocities due to technological barriers, such as the maximum sampling rate of high-speed cameras, drop of light source intensity with increasing pulsating frequency, and elevated cost of equipment. Post-processing \ac{TSS} strategies have thus been developed to increase artificially the temporal resolution of \ac{PIV} measurements. The existing techniques broadly fall into two main classes: model-free and model-based estimators. 

In the first category, an estimator is trained directly on the available data to increase the temporal resolution. This class includes estimators trained solely with snapshots (see e.g. Refs. \cite{druault2005use,legrand2011flow,kim2021unsupervised,mokhasi2009predictive,nguyen2010proper}), and methods aided by sensors. In the latter case, a mapping from a high-repetition-rate sensor to the full snapshots is built. The Extended Proper Orthogonal Decomposition method \citep[EPOD,][]{boree2003extended} has been widely used to correlate synchronized measurements with spatial resolution and temporally resolved point-wise measurements. 
\acused{EPOD}
\ac{EPOD} and its variants have been used in several applications, including jet flows \cite{tinney2008low}, wall-mounted obstacles \cite{hosseini2015sensor}, wake flows, \cite{bourgeois2013generalized,tu2013integration,discetti2018estimation,chen2022pressure}, and turbulent wall-bounded flow \cite{discetti2019characterization}. The most recent tendency is towards using deep learning nonlinear estimators  \citep[see, e.g.,][]{deng2019time,manohar2023temporal}. Since flow estimation is an ill-posed problem due to the limited number of sensors, all these methods leverage the principle that the flow evolves on low-dimensional attractors. This facilitates the mapping from the space spanned by the sensors to the space of the attractor. On the downside, this results in lower spatial resolution of the estimated field, sensitivity to noise, and the need to introduce intrusive probes in the measurement domain. 

Model-based methods leverage physical principles to fill the temporal gaps between the available snapshots. The earliest approaches exploited Taylor's hypothesis of frozen turbulence \cite{de2012pressure,scarano2012advection,beresh2016applications,jaunet2016pod,
laskari2016full,schneiders2018pressure,van2019pressure,vamsi2020reconstructing} to increase the temporal resolution using an advection model. Although appealing for its simplicity, the applicability of this approach is bound by the validity of Taylor's hypothesis, as well as by a dependence on the selection of the convection velocity. The performance of such methods was later observed to be improved by fusing the model predictions with sensor measurements via Kalman filtering \cite{wang2021model}. More advanced models, based on the equation of advection of vorticity, have been proposed to overcome this limitation. The Vortex-In-Cell technique \cite{schneiders2014time, schneiders2016dense} significantly improves \ac{TSS} if compared to simple advection models, although it requires complete 3D snapshots. This limits its applicability to cases in which volumetric flow-field measurements are feasible. Following the recent trends towards machine-learning modelling, some works proposed physics-informed methods to increase the temporal resolution of flow measurements e.g. \cite{soto2024complete,fukami2021machine,chen2021reconstructing,xie2018tempogan,liu2020deep}.

In this paper, we propose a technique to perform \ac{TSS} using a model extracted from data. The fundamental hypothesis is that the turbulent flow behaviour is driven by the so-called coherent structures \cite{jimenez2018coherent}, whose dynamics often evolve in low-dimensional attractors. Large-scale structures are the main actors in the momentum transport in turbulent flows. Low-order models can be extracted from the data and constructed to model the dynamics \cite{rowley2017model,mendez2023linear}.  For this purpose, we leverage \ac{POD} \cite{holmes2012turbulence} and the Galerkin modelling approach described in Ref.~ \cite{noack2005need}. Using a reduced-order POD representation we obtain a \ac{ROM} of the flow under investigation. Models based on the Galerkin projection of the Navier-Stokes equations over the subspace spanned by the \ac{POD} modes have been extensively studied for the identification of \acp{ROM} of dynamical systems across a wide range of applications. The number of  modes required to accurately capture the dominant flow dynamics varies depending on the specific flow configuration. For instance, several studies \cite{rajaee1994low,ukeiley2001examination,rowley2004model,wei2009low,deng2020low,sikroria2020application,deng2021galerkin,girfoglio2021pod,girfoglio2022pod,girfoglio2023hybrid} have demonstrated that a \ac{POD} basis truncated at under $10$ modes can suffice to capture the dominant dynamics in flows such as the wake of a cylinder, free shear layers, turbulent mixing layers, and compressible flows like cavity flows. On the other hand, for flows with more complex structure interactions, such as the transitional boundary layer analysed in Ref. \cite{rempfer2000low}, it is required $O(10)$ modes to capture the full dynamics of the fields. Furthermore, in cases dominated by multi-scale broadband interactions, the number of modes required for a complete description of the dynamics increases rapidly to the order of hundreds or even thousands of modes, as can be the case with the turbulent boundary layer or turbulent channel flow, as shown in Refs. \cite{aubry1988dynamics, omurtag1999low}. Thus, despite the effective low-dimensional representation of the dynamics for specific systems, such observations remark the limitations of the models based on Galerkin projections, where applicability becomes challenging when complexity increases and a large number of modes is necessary for the full reconstruction of the flow dynamics.

Recently, in Ref. \cite{asztalos2024galerkin}, a Galerkin spectral model was proposed for spectral analysis of time-super-sampled flow fields. The method proposed in this work exploits the same hypothesis of compact dynamics underlying the estimation from sensors and bridges the gap between model-free and model-based techniques for time resolution enhancement. The \ac{GM} is indeed on the one hand data-driven -\ac{POD} modes are tailored to the measured data- and on the other hand, it introduces a physical model based on the Navier-Stokes equations. The time history of the flow fields is reconstructed by integrating the \ac{ROM} in time, using as checkpoints the measured snapshots.

The paper first presents in Section \ref{sec:Galerkin model} a brief mathematical description of the proposed methodology for \ac{TSS} of \ac{NTR} \ac{PIV} measurements. Later, Section \ref{sec:datasets} describes the datasets used for the validation and testing of the model, thus first explaining a synthetic dataset of a fluidic pinball and an experimental \ac{PIV} dataset of a turbulent jet flow. The corresponding results from the reconstruction of time super-sampled fields are then discussed in Section \ref{sec:model validation}. The estimator's performance is assessed for increasing temporal separation between consecutive snapshots and different truncation levels. Finally, the main ideas and conclusions are summarised in Section \ref{sec:conclusion}.

\section{\label{sec:Galerkin model}Temporal resolution enhancement with Galerkin models}
The proposed model for time super-sampling of flow field measurements is based on a Galerkin projection of the Navier-Stokes equations on a suitable orthonormal basis, following the theoretical backbone of the work of Noack et al. \cite{noack2005need}. 
The selected basis is composed of a subset of the \ac{POD} spatial modes obtained from the ensemble of the \ac{NTR} velocity fields ($\mathbf{u}(\mathbf{x},t)$). Thanks to this projection, we describe the flow dynamics with a system of $N$ linearly dependent \acp{ODE}, one for each of the $N$ orthogonal \ac{POD} modes considered. The dynamical model is integrated in time to reconstruct the evolution of the flow field between the measured fields of the original dataset. 

\subsection{\label{sec:POD}Proper Orthogonal Decomposition}
\ac{POD} is a linear modal analysis tool widely used to identify modes containing the most energetic structures from a set of data \cite{berkooz1993proper}. Typically, \ac{POD} is employed for the spatio-temporal decomposition of velocity fields $\textbf{u}(\textbf{x},t)$. The corresponding modes are optimal in the least-square sense, i.e. they minimize the Frobenious norm of the flow field reconstruction. The modes are ordered by their energy content. 

\definecolor{darkred}{HTML}{aa0000}
\newcommand{\forwardint}{\raisebox{2pt}{\tikz{\draw[-,darkred,dashed,line width = 1pt](0,0) -- (5mm,0);}}}

\definecolor{darkblue}{HTML}{0055d4}
\newcommand{\backwardint}{\raisebox{2pt}{\tikz{\draw[-,darkblue,dashed,line width = 1pt](0,0) -- (5mm,0);}}}

\definecolor{grey}{HTML}{999999}
\newcommand{\weightint}{\raisebox{2pt}{\tikz{\draw[-,grey,solid,line width = 1pt](0,0) -- (5mm,0);}}}

\definecolor{darkgreen}{HTML}{2ca05a}
\newcommand{\darkgreendot}{\raisebox{0.6pt}{\tikz{\node[draw=black, scale=0.35, circle, fill=darkgreen](){};}}}

\begin{figure}[t]
    \centering
    \includegraphics[scale = 0.9]{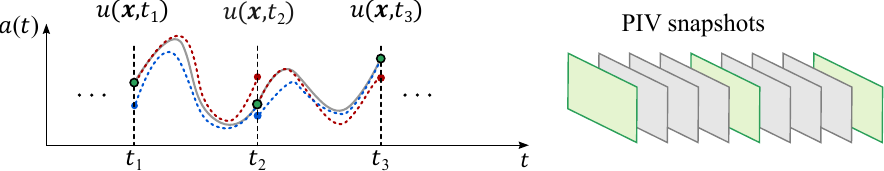}
    \caption{Schematics of the integration method for a generic temporal coefficient $a(t)$. Snapshots available in the \ac{NTR} dataset are depicted in green (\protect \darkgreendot). In red (\protect \forwardint) the results of the forward integration, and in blue (\protect \backwardint) the backward integration. In grey, the reconstructed snapshots and the weighted average solution between the forward and backwards integrations (\protect \weightint) .}
    \label{fig:integration}
\end{figure} 

Assume a set of $n_t$ measurements on $n_p$ points of the flow velocity is available. The fluctuating velocity fields $\mathbf{u}^{\prime}(\mathbf{x},t)$ are obtained through the \textit{Reynolds decomposition} ($\mathbf{u}(\mathbf{x},t) = \langle \mathbf{u} \rangle + \mathbf{u}^{\prime}(\mathbf{x},t)$), with $\langle \cdot \rangle$ indicating the ensemble average operator. The fluctuating velocity fields are rearranged into the so-called \textit{snapshot matrix} ($\mathbf{U}$ of size $n_p \times n_t$), where each temporal snapshot is reshaped in a column-wise fashion. The typical case of a \ac{NTR} \ac{PIV} dataset is that $n_t << n_p$, thus the \textit{snapshot method} \cite{sirovich1987turbulence} is employed for an efficient computation of the POD modes. Note that $n_p$ corresponds to the product of the total number of spatial points per snapshot times the dimension of the captured fields (i.e. $2$ for planar measurements or $3$ for stereoscopic or tomographic measurements). In this framework, with uniformly-spaced grids, the \ac{POD} can be obtained by computing an economy-size \ac{SVD}: 
\begin{equation}\label{eq:POD}
\mathbf{U} = \boldsymbol{\Phi}\;\boldsymbol{\Sigma}\;\boldsymbol{\Psi}^* = \boldsymbol{\Phi}\;\mathbf{A}
\end{equation}

In this decomposition, $\bm{\Phi} \in \mathbb{R}^{n_p\times n_t}$ and $\bm{\Psi} \in \mathbb{R}^{n_t\times n_t}$ are unitary matrices, composed respectively of orthogonal spatial modes $\bm{\phi}_i$ along its columns and temporal modes $\bm{\psi}_i$ along its rows. The diagonal matrix $\bm{\Sigma} \in \mathbb{R}^{n_t \times n_t}$ contains the singular values $\sigma_i$ of $\mathbf{U}$, i.e. the variance contribution of each mode to the snapshot matrix. Here, $\mathbf{A}=\boldsymbol{\Sigma}\;\boldsymbol{\Psi}^*$ is the matrix of temporal coefficients and $*$ denotes the Hermitian transpose. 

The velocity fields can be reconstructed as a linear combination of the $N=min(n_t, n_p)$ modes:
\begin{equation}
    \mathbf{u}(\mathbf{x},t) = \sum_{i = 0}^{N} \mathbf{a}_i(t) \;\boldsymbol{\phi}_i(\mathbf{x}), 
    \label{eq:reconstructionPOD}
\end{equation}

\noindent where index 0 refers to the mean velocity field such that $\boldsymbol{\phi}_0 = \frac{\langle\mathbf{u}\rangle}{\mid \langle\mathbf{u}\rangle \mid}$ and $a_0 = \mid \langle \mathbf{u} \rangle \mid$. The expansion coefficients are computed as $\mathbf{a}_i = (\mathbf{u} - \langle \mathbf{u}\rangle,\boldsymbol{\phi}_i)_{\Omega}$ \footnote{$(\mathbf{v},\mathbf{w})_{\Omega}=\int_{\Omega} dV \mathbf{v} \cdot \mathbf{w}$, inner product in the space of square-integrable vector field on the domain $\Omega$.}. Thus, a rank-$r$ \ac{LOR} can be obtained by truncating the sum in \eqref{eq:reconstructionPOD} to the first $r$ leading modes. This poses an inherent limit to the accuracy of the Galerkin model, since the dynamics of the least energetic scales are not captured.

\subsection{\label{sec:Galerkin projection}Galerkin modelling for time super-sampling}

In this work, we identify a low-dimensional empirical \ac{GM} \cite{noack2005need} based on the projection of the Navier-Stokes equations onto the low-dimensional \ac{POD} basis. This is the so-called \textit{Galerkin projection} \cite{holmes2012turbulence}, a mathematical technique to solve differential equations which transforms an infinite-dimensional \ac{PDE} into a finite set of \acp{ODE}. 

The dynamical system describing the evolution of the expansion coefficients $\mathbf{a}_i$ obtained from the Galerkin approximations is:
\begin{equation} \label{eq:Galerkin model}
    \frac{d}{dt}\mathbf{a}_i = \frac{1}{Re} \sum_{j = 0}^r l_{ij} \mathbf{a}_j + \sum_{j = 0}^r \sum_{k = 0}^r q_{ijk}\mathbf{a}_j \mathbf{a}_k + \mathbf{f}_i^{\pi}(\mathbf{A}) \quad \textrm{for} \quad i = 1,...,r.
\end{equation}
where $l_{ij} = (\phi_i,\Delta \phi_i)_{\Omega}$ are the viscous term coefficients, $q_{ijk} = (\phi_i,\nabla \cdot(\phi_j \phi_k))_{\Omega}$ the convective ones, and $\mathbf{f}_i^{\pi} = -(\phi_i,\nabla p)_{\Omega}$ is the pressure term. For a more detailed model description, the reader is referred to Ref.~\cite{noack2005need}. We assume in the following that only instantaneous velocity fields are available, in a scenario typical of standard \ac{PIV}. Under the conditions of insufficient temporal resolution, the pressure term in Eq. \ref{eq:Galerkin model} cannot be directly measured. Two exceptions hold in this context: availability to all boundary conditions would allow the pressure to be solved through the Poisson equation; alternatively, access to Multi-pulse \ac{PIV} systems \cite{hain2007fundamentals} would facilitate the estimation of instantaneous acceleration measurements and thus, the pressure gradient. Note that the assumption of null pressure term holds under certain conditions such as periodic boundary conditions and zero mean pressure gradient, namely `closed' flows \cite{holmes2012turbulence}, as with the case of the turbulent Couette flow \cite{moehlis2002models}. For the fluidic pinball, the pressure term typically vanishes for sufficiently large domains and is neglected in the reconstruction \cite{deng2020low}. For the jet flow, several studies neglect the contribution of the pressure term in the estimation of velocity fields \cite{townsend1976structure,alexander1953transport,liepmann1947investigations,schlichting2016boundary}, which typically accounts for $10\%$ of the total \cite{ito2021similarity}. The estimation of the pressure is out of the scope of this work, nonetheless, for the interest of the reader, some works for the pressure estimation are, i.e. \cite{van2013piv}.

For time super-sampling of \ac{NTR} \ac{PIV} snapshots, the model of the time coefficients is integrated in time to fill the gaps between the available snapshots of the dataset. A Runge-Kutta method of fifth order is used, integrated within the built-in function \texttt{solve\_ivp} from \texttt{Python}'s library \texttt{SciPy 1.13.1}, which uses an adaptive time step with a relative tolerance of $10^{-15}$ and an absolute tolerance of $10^{-10}$.

Model truncation, flow three-dimensionality and sensitivity to noise in the initial conditions are sources of error whose effects are amplified in the integration process. Furthermore, direct application of forward (in time) integration would result in discontinuities at each new initial condition. We propose a forward-backward integration to tackle these issues (see Fig. \ref{fig:integration}), using the snapshots of the \ac{NTR} dataset as initial conditions, similar to the work of \cite{vamsi2020reconstructing,wang2021model}. A weighted average between a forward and backward integration of snapshots is considered, with full weighting given to the initial condition snapshots to ensure compliance with the measured data. In analogy with Kalman filtering, this would correspond to a Kalman gain equal to 1. Namely, the first snapshot gets a weight equal to 1 in the forward integration and 0 in the backward, while the last snapshot gets 1 in the backward integration and 0 for the forward. A sigmoid function is used to weigh the integration results for the intermediate points. Therefore, reconstructed regions in-between \ac{NTR} snapshot pairs are independent among each other, depending only on the initial and final conditions. Note that the backward integration is performed by inverting the time axis, which may lead to numerical instabilities due to negative viscosity effects. Nonetheless, this has a minor effect if the viscous term of the equations has a small contribution compared to the convective term. In the test cases presented in this work, the viscous term is at least 2 orders of magnitude smaller than the convective term. The weighting process with the forward integration further reduces the risk of instability.

The corresponding velocity fields $\mathbf{u}_{GM}$ at the integrated instants are reconstructed by projecting back onto the $r$-rank POD basis obtained from the \ac{NTR} dataset:
\begin{equation}
    \mathbf{u}_{GM}(\mathbf{x},t^+) = \sum_{i = 0}^{r} \mathbf{a}_i(t^+) \;\boldsymbol{\varphi}_i(\mathbf{x}),
\end{equation}
\noindent with $t^+$ being an arbitrary time instant in the time span within consecutive acquisitions.

\begin{figure}[H]
    \centering
    \includegraphics[scale=0.8]{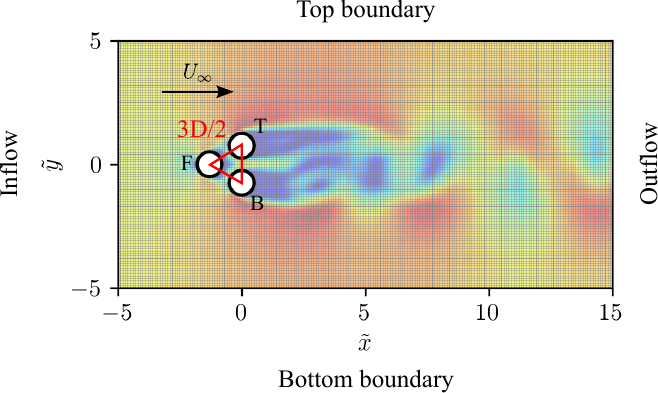}
    \caption{Schematics of the computational domain of the fluidic pinball indicating the boundaries, and the main flow and geometrical features. The uniform grid interpolated from the nonuniform \ac{DNS} grid is represented in grey. The circles represent the cylinders of the pinball, with the centres arranged in an equilateral triangle disposition of side $3D/2$. The cylinders are named as F (Front), T (Top), and B (Bottom). The direction and magnitude of the free-stream velocity $U_{\infty}$ is depicted with an arrow.}
    \label{fig:FP_Setup}
\end{figure}

\definecolor{lasergreen}{HTML}{00ff00}
\newcommand{\lasergreenline}{\raisebox{2pt}{\tikz{\draw[-,lasergreen,dashed,line width = 1pt](0,0) -- (5mm,0);}}}
\definecolor{red}{HTML}{ff0000}
\newcommand{\reddashed}{\raisebox{2pt}{\tikz{\draw[-,red,dashed,line width = 1pt](0,0) -- (5mm,0);}}}

\begin{figure}[H]
    \centering
    \includegraphics[scale=0.8]{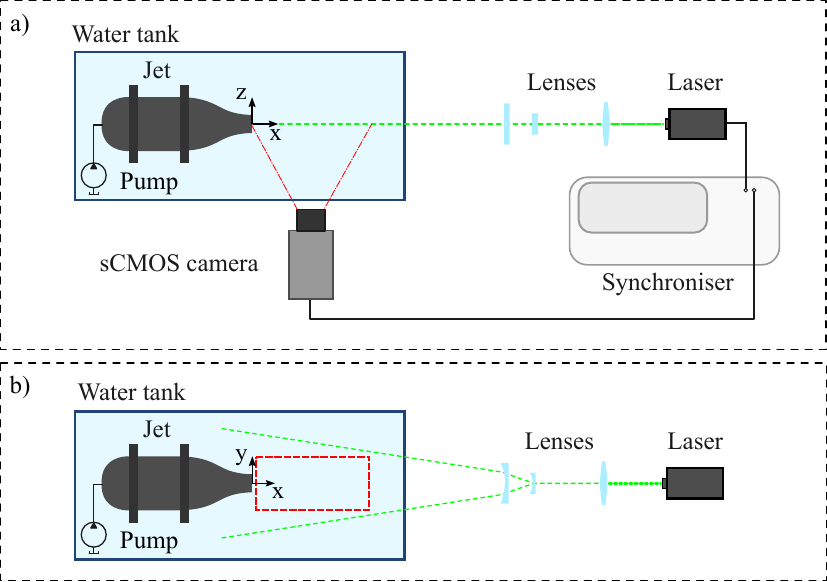}
    \caption{Schematic representation of the experimental PIV measurements of the turbulent jet flow. The plane of measurement is illuminated with a laser sheet (\protect\lasergreenline). The field of view of the PIV measurements in indicated with a box of red dashed lines (\protect\reddashed). Figure adapted from \cite{franceschelli2025assessment}, under license CC BY 4.0.}
    \label{fig:Jet_SetUp}
\end{figure}
\section{\label{sec:datasets}Datasets}
\subsection{\label{sec:fluidic pinball}Wake of fluidic pinball}
The first test case consists of a synthetic dataset from a 2D \ac{DNS} of the flow around and in the wake of a fluidic pinball \cite{deng2020low}. The simulation layout (see Fig. \ref{fig:FP_Setup}) features three cylinders with equal diameters $D$, each located in the vertex of an equilateral triangle of side length equal to $3D/2$, immersed in a flow with uniform velocity $U_{\infty}$. The downstream cylinders are located at $x/D = 0$ and centred with respect to the $y$ axis, corresponding to the crosswise direction. \ac{DNS} data are computed at $Re = 130$, corresponding to a symmetric chaotic regime \cite{deng2020low}. 
The simulation is computed with non-slip boundary conditions on the surface of the cylinders, free flow across the inflow, top, and bottom boundaries, as well as stress-free conditions in the outflow region. Zero-mean Gaussian random noise with $1\%$ of standard deviation is added to each snapshot to evaluate the effect of noise.

A total of $10000$ \ac{TR} snapshots are generated, with time separation $\Delta\Tilde{t} = 0.1$, where the time variable is normalised with the convective time $\Tilde{t} = t D/U_{\infty}$. The flow velocity fields are interpolated on a Cartesian grid extending between $\Tilde{x} = x/D \in [-5,15]$ in the streamwise direction, and $\Tilde{y} = y/D \in [-5,5]$ in the crosswise direction, with a grid spacing of $d\Tilde{x} = d\Tilde{y} = 0.1$. This corresponds to a resolution of $10\si{px/D}$ and a total number of grid points of $N_p = N_x \times N_y = 20301$.

\definecolor{tabgreen}{HTML}{2ca02c}
\definecolor{tabblue}{HTML}{1f77b4}
\definecolor{tabred}{HTML}{d62728}
\definecolor{taborange}{HTML}{FF7F0E}
\definecolor{myblue}{HTML}{069af3}

\newcommand{\greendot}{\raisebox{0.6pt}{\tikz{\node[draw=tabgreen, scale=0.3, circle, fill=tabgreen](){};}}}
\newcommand{\reddot}{\raisebox{0.6pt}{\tikz{\node[draw=tabred, scale=0.3, circle, fill=tabred](){};}}}
\newcommand{\bluedot}{\raisebox{0.6pt}{\tikz{\node[draw=tabblue, scale=0.3, circle, fill=tabblue](){};}}}
\newcommand{\orangedot}{\raisebox{0.6pt}{\tikz{\node[draw=taborange, scale=0.3, circle, fill=taborange](){};}}}

\newcommand{\blackline}{\raisebox{2pt}{\tikz{\draw[-,black,solid,line width = 1pt](0,0) -- (5mm,0);}}}

\newcommand{\myblueline}{\raisebox{2pt}{\tikz{\draw[-,myblue,solid,line width = 1pt](0,0) -- (5mm,0);}}}

\begin{figure}[t]
    \centering
    \includegraphics[trim={0cm 0cm 0cm 1cm},clip, scale = 0.9]{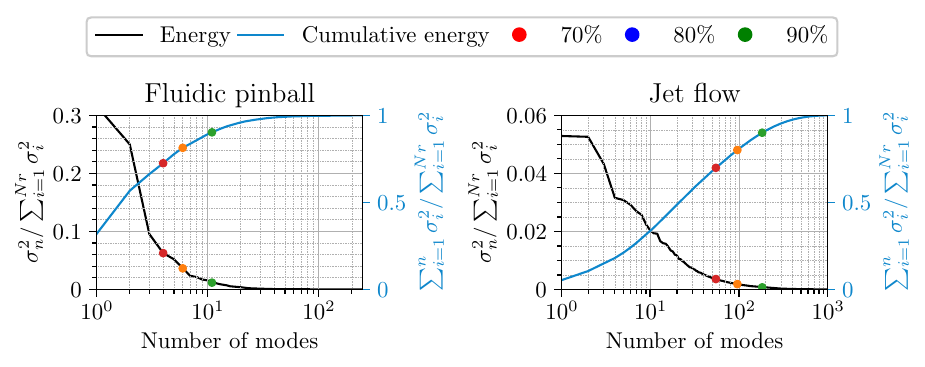}
    \caption{Energy (\protect\blackline) and cumulative energy (\protect\myblueline) distribution for increasing number of modes. On the left, the fluidic pinball and on the right, the jet flow. The rank truncations used for the analysis are represented with dots, respectively $70\%$ (\protect\reddot), $80\%$ (\protect\orangedot), and $90\%$ (\protect\greendot) of the total energy.}
    \label{fig:modes energy}
\end{figure}

\subsection{\label{sec:jet flow}PIV experiment of a jet flow}

The method is then evaluated on experimental planar \ac{PIV} measurements of a water jet flow (see Fig. \ref{fig:Jet_SetUp}). The objective of this test case is to assess the performance of our \ac{TSS} technique with data corrupted by noise and stronger truncation effects due to the higher rank of the underlying flow and the 3D effects that cannot be captured with the 2D \ac{PIV} data. 

The experiments are performed in the water tank facility of the Experimental Aerodynamics and Propulsion Lab group at Universidad Carlos III de Madrid. The tank has dimensions $80\times 60 \times 40 \si{cm^3}$. The circular jet nozzle has an exit diameter of $D = 0.03\si{m}$. It is operated at a bulk velocity of approximately $U_b \approx 0.11\si{m/s}$, yielding a Reynolds number $Re \approx 3300$ based on $U_b$ and $D$. The flow is illuminated on the symmetry plane of the jet with a \textit{LaserTree LT-40W-AA} pulse-width modulated laser with a power of $5\si{W}$. The light source is shaped with a spherical converging and a cylindric diverging lens into a laser sheet of approximately $1\si{mm}$ thickness at the exit of the nozzle. The flow is seeded with polyamide particles of $56\si{\micro m}$ diameter. \ac{TR} planar \ac{PIV} snapshots are captured with an \textit{Andor Zyla sCMOS camera} of a $5.5\si{megapixel}$ sensor ($2160\times 2560 \si{px}$, pixel size of $6.5\si{\micro m}$). The image is captured with a spatial resolution of $6.53\si{px/mm}$ in a domain spanning from the nozzle exit at $\Tilde{x} = x/D=0$ to $\Tilde{x} = 8$ downstream and $\Tilde{y} = y/D \in [-1.3,1.3]$ in the crosswise direction, being this axis centred in the nozzle axis of symmetry. The reader is referred to \cite{franceschelli2025assessment} for a more detailed description of the experimental setup of the jet flow. 

A \ac{TR} sequence of $60000$ snapshots is captured with a time separation of $\Delta t = 0.0011\si{s}$. This is equivalent to $0.04$ convective times, being the convective time computed using $U_b$ and $D$. The \ac{PIV} images are processed with PaIRS\footnote{Open software for processing of PIV images developed by the Experimental Thermo Fluid Dynamics group of the Università Federico II di Napoli.} \cite{astarita2005analysis,Paolillo2024}, with a multi-grid/multi-pass \cite{soria1996investigation} image deformation algorithm \cite{scarano2001iterative}. The final interrogation windows are of $32\times 32\si{pixels}$, with $75\%$ overlap. The velocity fields are then post-processed with a $2^{nd}$-order polynomial Savintzky-Golay filter with a $5\times 5$ kernel in space and $5$ snapshots in time. A Gaussian filter of standard deviation $\sigma = 1$ and a kernel size of $7$ is applied to enhance the smoothing of temporal data.


\section{\label{sec:model validation}Validation}
The assessment of the Galerkin time super-sampling procedure is verified with the two \ac{TR} measurements described in the previous section. The datasets are first preprocessed to obtain downsampled fields. For the fluidic pinball, the snapshots to compute the Galerkin model are directly obtained from the full sequence by skipping snapshots. Consequently, the number of snapshots used to train the basis changes with the downsampling factor. For the jet flow, instead, a part of the sequence is retained in full to ensure that a fixed number of $1000$ snapshots is used. In all cases, the snapshots to be reconstructed by the Galerkin model are never used for training. The discarded snapshots are employed as ground truth to evaluate the super-sampling capabilities of the model. In the following, we analyze the effects of mode truncation and insufficient temporal resolution.

The \ac{DNS} and \ac{TR PIV} data are downsampled by a factor of $S_r$, i.e. we consider one snapshot every $S_r$ snapshots. The retained snapshots after downsampling are used to compute the \ac{GM} coefficients and as initial conditions for time integration. The full \ac{TR} dataset is only considered for the error analysis of the reconstructed fields. The parametric study is performed for $S_r = 10, 20,$ and $40$. The corresponding time steps between snapshots result in $\Delta \tilde{t}_s = 1, 2,$ and $4$ for the fluidic pinball, and $\Delta \tilde{t}_s = 0.4, 0.8,$ and $1.6$ for the jet flow. 

Furthermore, the effect of the amount of energy described by the \ac{ROM} is assessed. We show the results for \ac{ROM} accounting for $70\%$, $80\%$, and $90\%$ of the flow variance. Such thresholds in reconstructed energy have been manually selected as a compromise between a representative basis and a moderate computational effort when dealing with its derived ODE system. Then, comparing the performance of the model with DNS data and with real experimental measurements, we assess the limitations introduced by highly turbulent, nonlinear, and three-dimensional effects which can not be described with a 2D \ac{ROM}. 

A parametric study is performed to analyse the impact of the super-sampling factor $S_r$ and the reconstructed energy percentage on the accuracy of the estimated flow fields (see Fig. \ref{fig:modes energy}). As a baseline for comparison, we use a direct interpolation of the \ac{POD} time coefficients with a cubic spline. A visual comparison of the reconstructed velocity fields is also provided for the most relevant cases. In particular, we show one sample snapshot located halfway between retained snapshots. The rationale is that, in this case, the error introduced by the model approximation would be statistically maximum. Moreover, a spectral analysis is performed to assess the reconstruction capabilities of the model in terms of frequency content. A one-sided \ac{PSD} is computed according to:

\begin{equation}\label{eq:PSD}
    PSD = 2\cdot \frac{\mid\mid X \mid\mid ^2 }{N_s\cdot f_s},
\end{equation}

\noindent where $\mid\mid X \mid\mid$ is the magnitude of the Fourier transform coefficients of the signal -in our case, streamwise or crosswise fluctuating velocity components-, $N_s$ is the number of samples per window, and $f_s$ is the sampling frequency. The \ac{PSD} is computed numerically at representative points using Welch's method \cite{Welch1967}, applying a Hanning windowing.

The accuracy of time super-sampling of \ac{NTR} snapshots is estimated in terms of the mean cosine similarity ($S_c$) between the velocity fields reconstructed by interpolation or with the \ac{GM}($\mathbf{u}_{recon}$), and the reference velocity fields ($\mathbf{u}_{ref}$), computed in terms of the time separation from the measured snapshots according to Eq. \ref{eq:cosine similarity}. The original \ac{TR} snapshots are used as reference to assess the reconstruction error due to the truncation, which will be present in all cases. Therefore, the \ac{LOR} is used as the benchmark to evaluate the reconstruction performance with the \ac{GM} and the interpolation.

For the sake of comparison, the reconstructed snapshots are extracted with a $\Delta \tilde{t}$ equal to the time separation between the \ac{TR} snapshots of the reference dataset. 
\begin{equation} \label{eq:cosine similarity}
    S_c = \frac{\mathbf{u}_{ref}^{\prime} \cdot \mathbf{u}_{recon}^{\prime}}{\lVert \mathbf{u}_{ref}^{\prime} \rVert \cdot \lVert \mathbf{u}_{recon}^{\prime} \rVert }.
\end{equation}

The cosine similarity allows for the evaluation of the alignment of vectors between the reference velocity fields and the reconstructed fields. Moreover, the mean spatial distribution of the reconstruction error is analysed in terms of the mean square error ($MSE$) to identify the regions where the reconstruction is more prone to performance loss. This $MSE$ is computed according to:
\begin{equation} \label{eq:MSE}
    MSE_{u^{\prime}, v^{\prime}} = \frac{1}{n_t \; (\sigma_{u^{\prime},ref} ^2 + \sigma_{v^{\prime},ref}^2)} \sum_{i = 1}^{n_t} \bigl[(u_{ref,i}^{\prime} - u_{recon,i}^{\prime})^2 + (v_{ref,i}^{\prime} - v_{recon,i}^{\prime})^2 \bigr],
\end{equation}

\noindent
where $\sigma_{u^{\prime},TR}$ and $\sigma_{v^{\prime},TR}$ are the standard deviation of the reference velocity fluctuations and $u^{\prime}_{GM,i}$ and $v^{\prime}_{GM,i}$ are estimated velocity fluctuation components in the streamwise and spanwise direction at time step $i$.

The error of the estimated temporal coefficients ($a_{recon}$) per mode is analysed in terms of the root mean square error $\delta_{\epsilon}$ with respect to the reference $a_{ref}$ as:
\begin{equation} \label{eq:delta_eps}
    \delta_{\epsilon,i} = \frac{1}{\sigma_{a,ref,i}} \Bigg[\frac{1}{n_t}\sum_{j = 0}^{n_t} (a_{ref,i}(t_j) - a_{recon,i}(t_j))^2 \Bigg]^{1/2} \quad \textrm{for} \quad i = 1,..., r.
\end{equation}

With this, the quality of the reconstructed modes is evaluated in terms of the temporal separation between samples and the energy truncation levels.

\definecolor{myred}{HTML}{ff6f52}
\definecolor{myblue}{HTML}{069af3}

\newcommand{\redline}{\raisebox{2pt}{\tikz{\draw[-,myred,line width = 1pt](0,0) -- (5mm,0);}}}

\newcommand{\blueline}{\raisebox{2pt}{\tikz{\draw[-,myblue,densely dashdotted, line width = 1pt](0,0) -- (5mm,0);}}}

\begin{figure}[H]
\centering
    {\begin{overpic}[trim={0cm 0cm 0cm 1cm}, clip, scale=0.8]{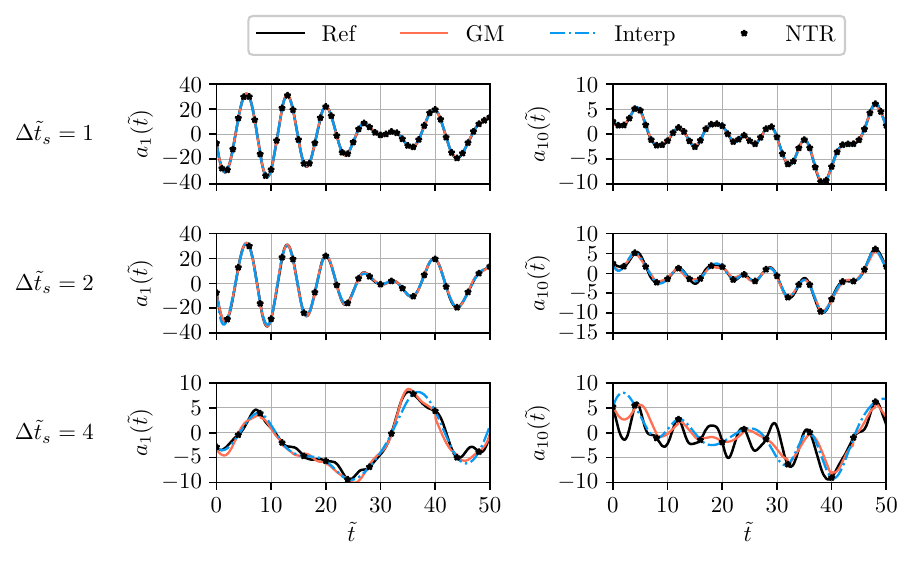}
    \end{overpic}
    }
    \caption{First and last temporal coefficients of the truncated fluidic pinball basis. On the left the first mode temporal coefficients, and on the right column the final mode. From top to bottom downsampling of $\Delta \Tilde{t}_s = 1, 2, 4$. The LOR is depicted with a black line (\protect\blackline) and the NTR dataset is marked with stars ($\boldsymbol{\ast}$). In orange (\protect\redline), the reconstructed snapshots with the \ac{GM}. In blue (\protect\blueline), the reconstruction with the mode interpolator.}
    \label{fig:FP_modes}
\end{figure}

\definecolor{myorange}{HTML}{f4320c}
\definecolor{myblue}{HTML}{3778bf}

\newcommand{\myline}{\raisebox{2pt}{\tikz{\draw[-,myblue, line width = 1pt](0,0) -- (5mm,0);}}}

\newcommand{\orangeline}{\raisebox{2pt}{\tikz{\draw[-,myorange, line width = 1pt](0,0) -- (5mm,0);}}}

\begin{figure}[H]
\centering
    {\begin{overpic}[trim={0cm 0cm 0cm 1cm}, clip, scale=0.8]{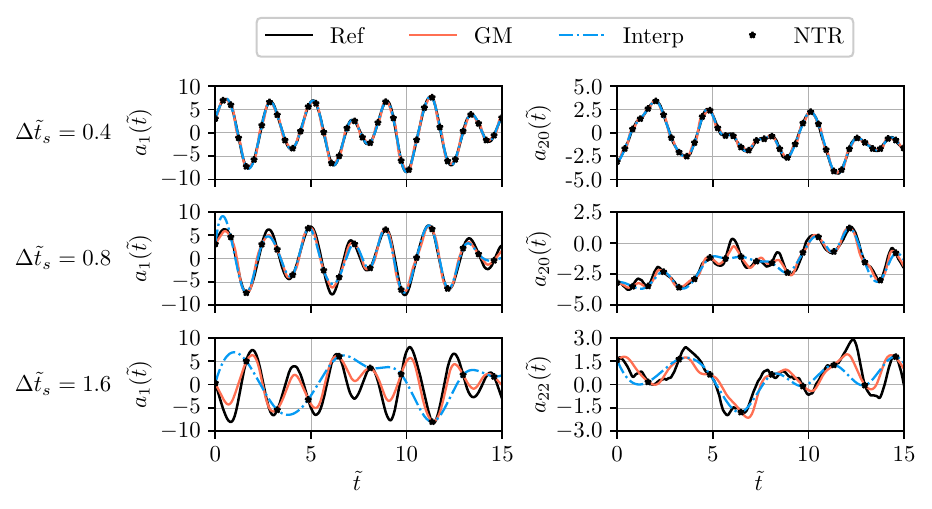}
    \end{overpic}
    }
    \caption{First and $1\%$ energy mode temporal coefficients of the truncated jet flow basis. On the left the first mode temporal coefficients, and on the right column the $1\%$ energy mode mode. From top to bottom downsampling of $\Delta \Tilde{t}_s = 0.4, 0.8, 1.6$. The \ac{LOR} is depicted with a black line (\protect\blackline) and the \ac{NTR} dataset is marked with stars ($\boldsymbol{\ast}$). In orange (\protect\redline), the reconstructed snapshots with the \ac{GM}. In blue (\protect\blueline), the reconstruction with the mode interpolator. }
    \label{fig:Jet_modes}
\end{figure}

\subsection{Analysis of super-sampling factor}
This section presents the results obtained for different time super-sampling factors of the NTR sequence. The reconstructed velocity fields are compared to the original TR velocity fields, thus in order to ensure the same time spacing between snapshots, the time super-sampling is analysed in terms of the downsampling of the TR dataset. Note that the same parameter value for each flow test case will have a different associated time separation between NTR snapshots. 

\begin{figure}[H]
    \centering
    {\begin{overpic}[scale=0.8,trim={0 0.2cm 0 0},clip]{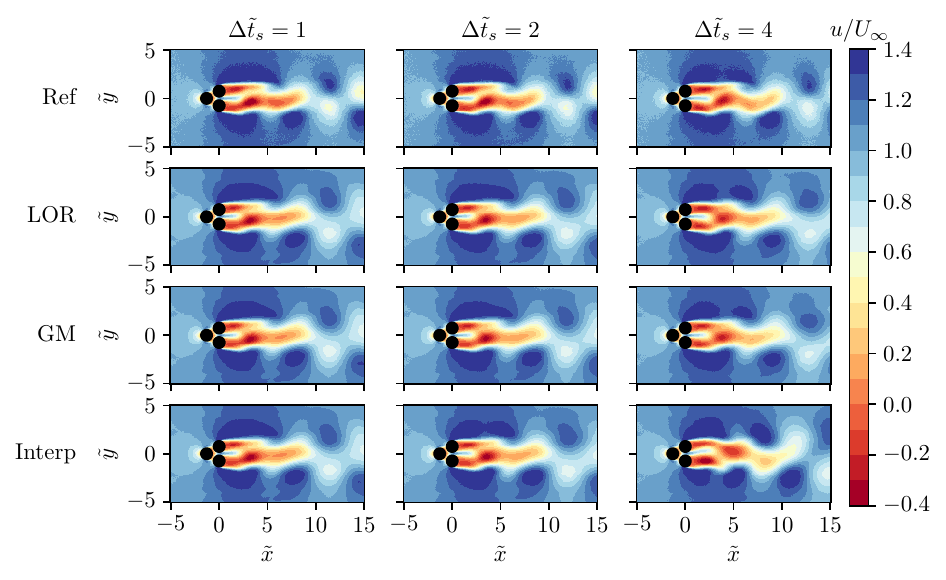}
    \end{overpic}
    }
    \caption{Reconstruction of streamwise velocity component of in-sample snapshots with $90\%$ of the total energy of the fluidic pinball. The snapshot corresponds to the midway snapshot between two consecutive snapshots of the \ac{NTR} sequence. From left to right, reconstructed snapshot for downsampling of $\Delta \Tilde{t}_s = 1, 2, 4$. From top to bottom, reference \ac{TR} field, \ac{LOR}, \ac{GM} reconstruction and interpolated reconstruction.}
    \label{fig:FP_snaps}
\end{figure}

\begin{figure}[H]
    \centering
    {\begin{overpic}[scale=0.8,trim={0 0.2cm 0 0},clip]{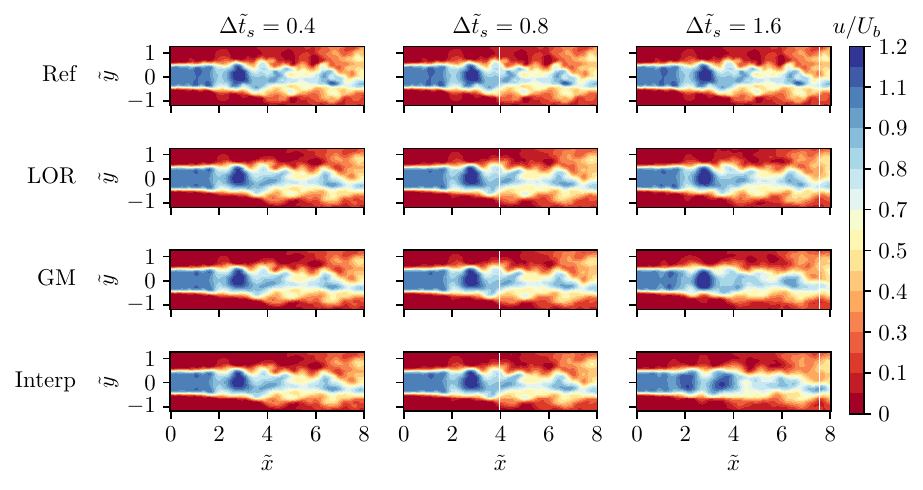}
    \end{overpic}
    }
    \caption{Reconstruction of streamwise velocity component of in-sample snapshots with $90\%$ of the total energy of the jet flow. The snapshot corresponds to the midway snapshot between two consecutive snapshots of the \ac{NTR} sequence. From left to right, reconstructed snapshot for downsampling of $\Delta \Tilde{t}_s = 0.4, 0.8, 1.6$. From top to bottom, reference \ac{TR} field, \ac{LOR},\ac{GM} reconstruction and interpolated reconstruction.}
    \label{fig:Jet_snaps}
\end{figure}

Three subsampling factors are considered ($S_r = 10,\;20,\;40$) with a constant energy reconstruction of $90\%$, selected to minimize the truncation error of the reconstructions. The ability to accurately enhance the temporal resolution is analyzed in terms of the time history of the temporal coefficients $\mathbf{a}_i$ (defined in Section \ref{sec:Galerkin projection}) (see Figs. \ref{fig:FP_modes} and \ref{fig:Jet_modes}). The most energetic mode and a smaller energy mode are provided to respectively assess the performance on the most relevant contribution to the dynamics of the flow, as well as the smallest reconstructed structure. The smaller energy mode will either consist of the last mode of the truncated basis or the last one to represent more than $1\%$ of its energy. At a small downsampling $S_r = 10$ both the interpolator and the \ac{GM} are able to reconstruct the snapshots with small errors. At $S_r = 20$, the performance of the interpolator starts degrading with respect to the Galerkin model, as the reconstruction of the higher order modes starts differing from the LOR. Finally, at a large downsampling of $S_r = 40$ the interpolator is inevitably affected by aliasing, while the  \ac{GM} still provides an acceptably accurate reconstruction. Note that the reconstruction of the least energetic mode is noticeably affected when increasing the temporal separation of the snapshots, which is induced by the truncation of the reconstruction as smaller scales are typically more sensitive to the interaction among them, thus retrieving a lower performance in the reconstruction when lower energy modes are truncated. 

\begin{figure}[t]
    \centering
    \includegraphics[scale=0.9,trim={0 0.2cm 0 0},clip]{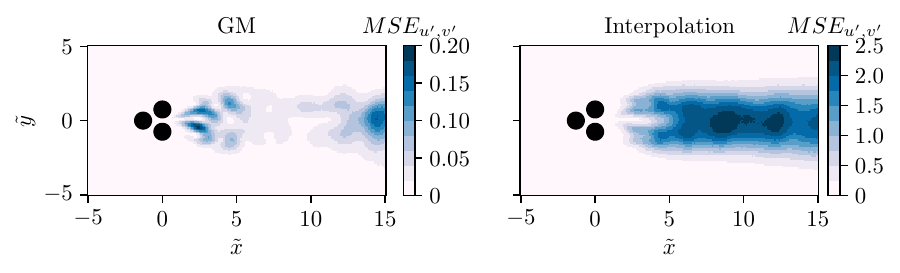}
    \caption{Mean square error map of the reconstructed fluctuation velocity fields for the fluidic pinball with a downsampling factor of $\Delta \Tilde{t}_s = 4$ and $90\%$ energy. On the left, error of the \ac{GM}. On the right, error of the interpolator.}
    \label{fig:FP_error}
\end{figure}

\begin{figure}[t]
    \centering
    \includegraphics[scale=0.9,trim={0 0.2cm 0 0},clip]{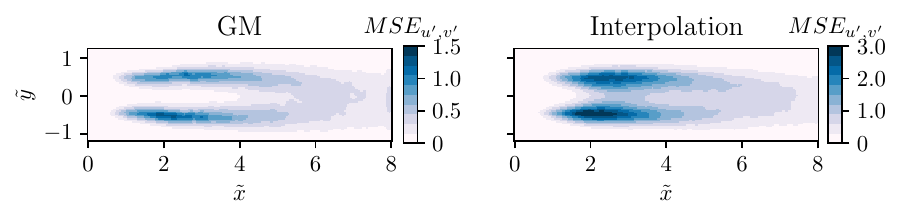}
    \caption{Mean square error map of the reconstructed fluctuation velocity fields for the jet flow with a downsampling factor of $\Delta \Tilde{t}_s = 1.6$ and $90\%$ energy. On the left, error of the \ac{GM}. On the right, error of the interpolator.}
    \label{fig:Jet_error}
\end{figure}

Furthermore, it can be remarked that the maximum error of the reconstruction is found at the mid snapshots between consecutive ones used as initial conditions, due to the propagation of numerical errors during the time marching integration. For small downsampling values, corresponding to sampling frequencies within the Nyquist limit, both the interpolator and the \ac{GM} have similar values of the reconstruction error.

Figs. \ref{fig:FP_snaps} and \ref{fig:Jet_snaps} show a snapshot located halfway between two snapshots of the NTR dataset used as initial conditions (referred to as \textit{mid snapshot}), where the error is statistically maximum. Each column represents the mid snapshot for increasing downsampling factors, and the rows depict the reference snapshots (\ac{TR} snapshot), the \ac{LOR}, the interpolation, and the Galerkin reconstruction, from top to bottom. Both datasets and reconstruction strategies present similar performances for downsamplings of $S_r = 10, 20$, thus correctly reconstructing the dynamics of the flow compared to the \ac{LOR}. However, for the maximum downsampling, the interpolator fails to reconstruct the structures effectively, as aliasing results in an out-of-phase representation leading to wrong flow structures when compared to the LOR. On the other hand, the \ac{GM} accurately reconstructs the larger-scale structures driving the main dynamics of the flow. The smaller structures, as anticipated previously, are more sensitive to truncation and retrieve a reconstruction more prone to failure. 

The time-averaged error maps between the LOR and the reconstructions for the most critical downsampling factor allow identifying the pivotal areas where the models are more prone to failure during the reconstruction of their dynamics (see Figs. \ref{fig:FP_error} and \ref{fig:Jet_error}). Thus, the reconstruction error is larger in the regions of vortex breakdown and intense intermittency. In the fluidic pinball, the error is one order of magnitude smaller than for the interpolation, with peak errors located in regions of the wake, especially far downstream where finer vortical structures are encountered. The interpolation fails to effectively reconstruct the turbulent wake, having a maximum error downstream around the centre line where most of the aliased dynamics occur. 

Similarly, the jet flow spatial error is most significant along the shear layer of the jet, with the \ac{GM} presenting half of the maximum error of the interpolation within a more delimited region. Since the jet flow consists of a highly turbulent flow with a three-dimensional interaction of structures and energy distributed along a wide range of scales, when performing the low-order reconstruction the dynamics of the smaller structures is lost, thus increasing the errors in the shear layer of the jet. 

\newcommand{\reddotted}{\raisebox{2pt}{\tikz{\draw[-,red,dotted, line width = 1.1pt](0,0) -- (5mm,0);}}}
\newcommand{\redsolidline}{\raisebox{2pt}{\tikz{\draw[-,red,solid, line width = 1pt](0,0) -- (5mm,0);}}}

\begin{figure}[t]
    \centering
    {\begin{overpic}[trim={0cm 0cm 0cm 1.3cm}, clip, scale=0.9]{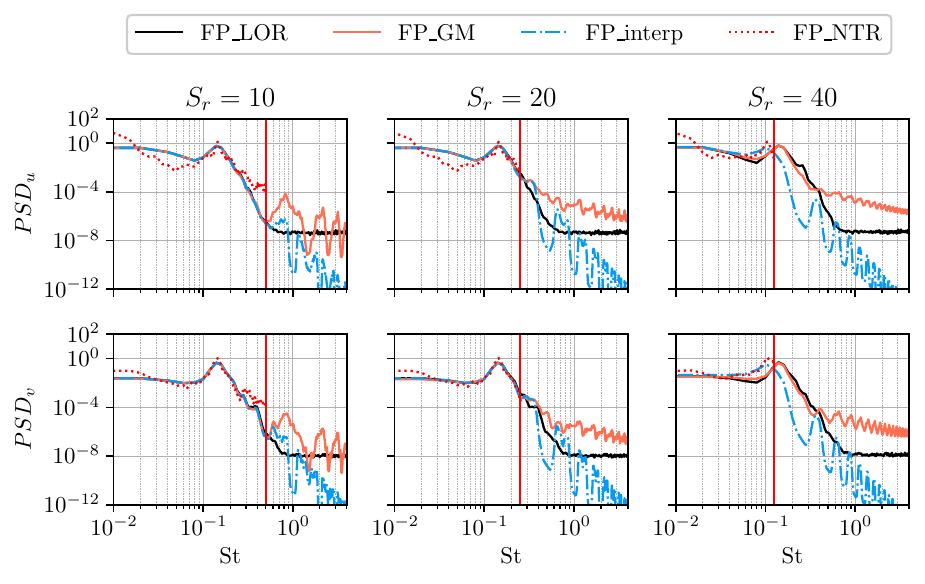}
    \put(-3,50) {\parbox{12mm}{\centering a)}}
    \end{overpic}
    }
    \vspace{0.5cm}
    {\begin{overpic}[trim={0cm 0.3cm 0cm 1.2cm}, clip, scale=0.9]{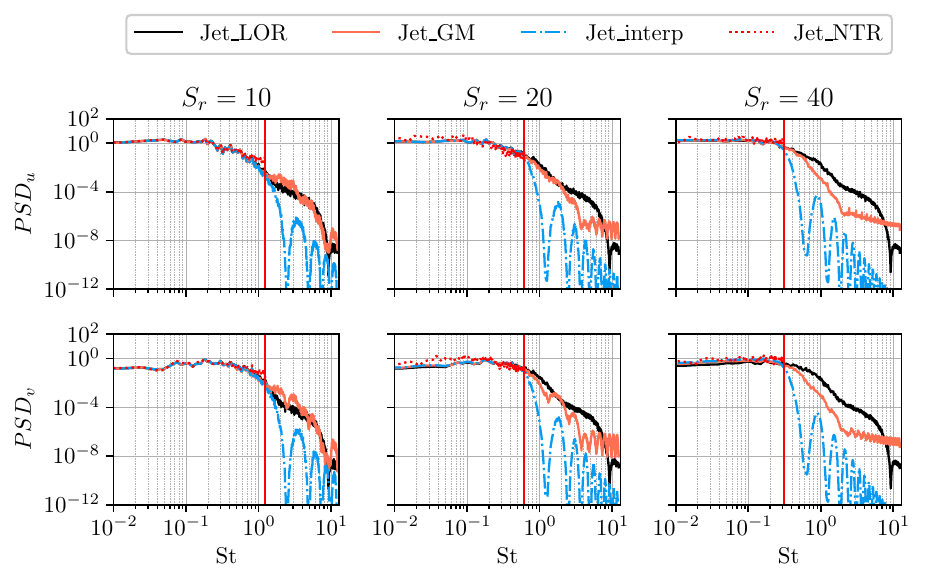}
    \put(-3,50) {\parbox{12mm}{\centering b)}}
    \end{overpic}
    }
    \caption{PSD of the results of the LOR, GM, interpolation and NTR dataset for different subsampling factors. The LOR is depicted in black (\protect \blackline), the NTR dataset with red dashed line (\protect \reddotted), the reconstructed snapshots with the GM in orange (\protect\redline), and the reconstruction with the mode interpolator in blue (\protect\blueline). The Nyquist limit is indicated by a red line (\protect \redsolidline). Subfigure a) shows the PSD of the streamwise (first row) and crosswise (second row) velocity at a point in the wake of the top cylinder, located at $x=2.5D, y=0.7D$. Subfigure b) shows the PSD of the streamwise (first row) and crosswise (second row) velocity at the lip-line in the developed region of the jet, in the point at $x=7D, y=-0.4D$.} 
    \label{fig:PSD analysis}
\end{figure}

The \ac{PSD} of the reconstructed fields is evaluated for the reconstructions with $90\%$ energy to assess the capability of the method to capture the spectral content of the flow. Additionally, a better insight to the total energy reconstruction can be assessed. 

The \ac{PSD} of the velocity fluctuations in the wake of the fluidic pinball shows a peak at $St = 0.14$, corresponding to the shedding frequency. This peak is correctly captured with the cases of $S_r = 10$ and $S_r = 20$ also by the \ac{NTR} data. However, at $S_r = 40$, the shedding frequency is located beyond the Nyquist limit of the \ac{NTR} data. The peak thus appears at a lower Strouhal number due to aliasing of the signal. Comparing the spectrum of the \ac{GM} and interpolated reconstruction for the latter case, it can be noted that the \ac{GM} is able to recover correctly the shedding frequency that was lost in the \ac{NTR} dataset. The interpolation does not recover this frequency.

It can be observed that the \ac{GM} presents higher noise in the reconstruction, especially when increasing the super-sampling factor. The ripples appearing in the reconstruction at higher frequencies are generated by the sigmoid function used in the forward-backward integration for the \ac{GM}, and the cubic-spline interpolation. On the other hand, a significant damping of the energy at high frequency is observed in the data reconstructed by interpolation of the \ac{POD} time coefficients.

\definecolor{midred}{HTML}{f46d43}
\definecolor{midblue}{HTML}{74add1}

\newcommand{\midredline}{\raisebox{2pt}{\tikz{\draw[-,midred,solid,line width = 1pt](0,0) -- (5mm,0);}}}

\newcommand{\midblueline}{\raisebox{2pt}{\tikz{\draw[-,midblue,solid,line width = 1pt](0,0) -- (5mm,0);}}}

\begin{figure}[t]
    \centering

    {\begin{overpic}[trim={0cm 0cm 0cm 1cm}, clip, scale=0.9]{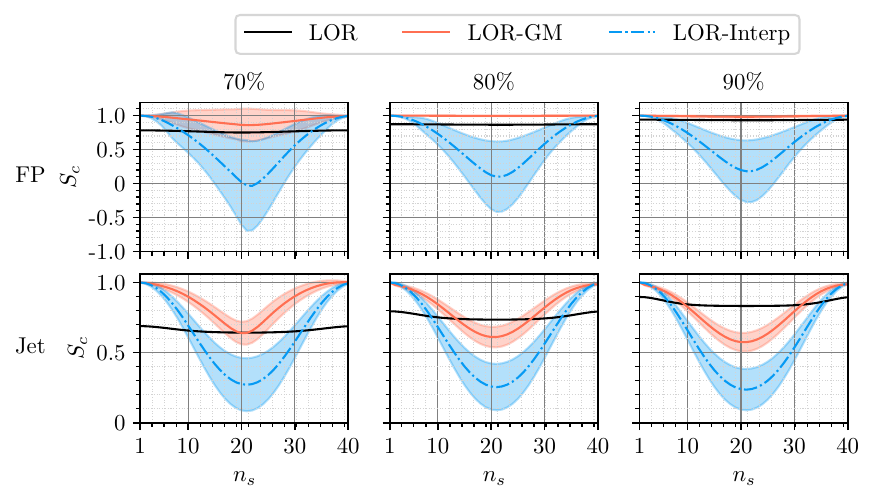}
    \end{overpic}
    }
    \caption{ Ensemble-average in time of the mean flow cosine similarity $S_c$ as a function of the separation from the measured snapshots, with $S_r=40$. The black line (\protect\blackline) depicts the $S_c$ of the \ac{LOR} with respect to the reference fields. The $S_c$ of the interpolation and the reconstruction with the \ac{GM} are represented in blue (\protect\blueline) and orange (\protect\redline), respectively. The corresponding standard deviation is depicted with the shading area.}
    \label{fig:cosine similarity}
\end{figure}


\begin{figure}[t]
    \centering
    \includegraphics[scale = 0.9]{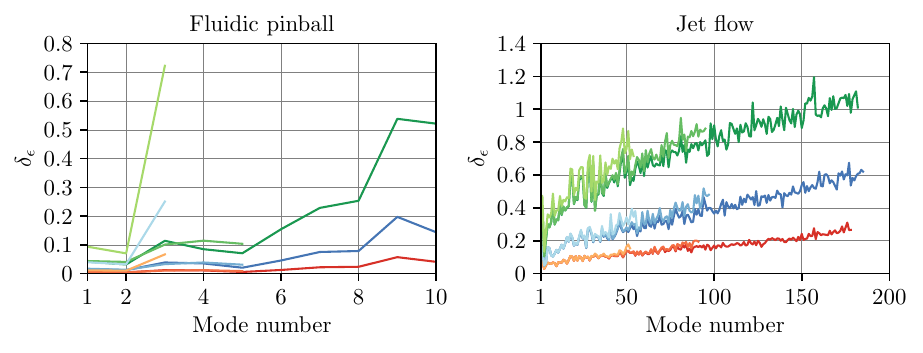}
    \caption{Root mean square error $\delta_{\epsilon}$ as a function of the mode number ($r$). On the left the fluidic pinball and on the right the jet flow. In green, the dataset with a subsampling $S_r = 40$, blue subsampling $S_r = 20$, and in red subsampling $S_r = 10$. The energy reconstructed is represented with the color intensity, from lighter to darker shades, $70\%$, $80\%$, and $90\%$, respectively.}
    \label{fig:error-modes}
\end{figure}

The \ac{PSD} of the jet, on the other hand, clearly shows the capability to extend the range of frequency with respect to the \ac{NTR} measurements and the interpolated sequence. The interpolation presents a rapid decay in the \ac{PSD} at the limit of the \ac{PSD} of the \ac{NTR} dataset. On the other hand, the \ac{GM} recovers the spectrum in a wider range of frequencies compared to the interpolation. The spectrum is observed to depart earlier from the reference one at higher supersampling factors, as the small-scale dynamics are not fully captured.

\newcommand{\mybline}{\raisebox{2pt}{\tikz{\draw[-,myblue, line width = 1pt](0,0) -- (5mm,0);}}}
\newcommand{\myorangeline}{\raisebox{2pt}{\tikz{\draw[-,myorange, densely dashdotted, line width = 1pt](0,0) -- (5mm,0);}}}

\begin{figure}[t]
    \centering
    \includegraphics[scale=0.9,trim={0cm, 0.2cm, 0cm, 1cm},clip]{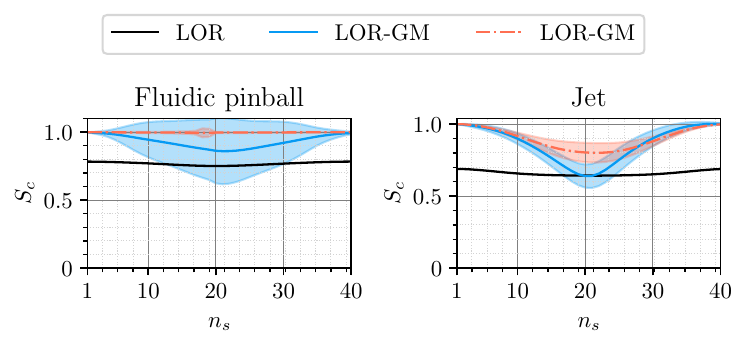}
    \includegraphics[scale=0.9,trim={0cm, 0cm, 0cm, 0.2cm},clip]{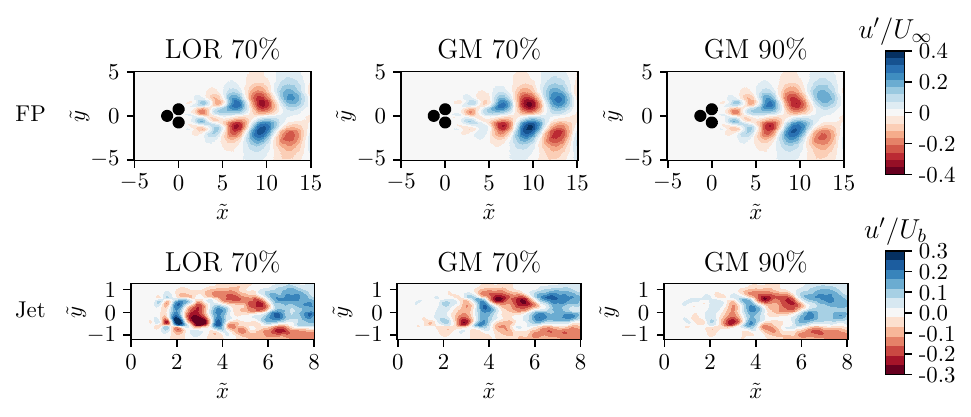}
    \caption{Ensemble-average in time of the mean flow cosine similarity and reconstruction of mid snapshot streamwise velocity fluctuation with $70\%$ of the energy of the systems truncated at $70\%$ and $90\%$. The first column depicts the mean cosine similarity for the reconstruction based on the truncated model at $70\%$ (\protect\mybline) and $90\%$ (\protect\myorangeline), with the standard deviation represented with the shaded area.}
    \label{fig:FOM70}
\end{figure}
The effect of energy reconstruction for the worst downsampling scenario $S_r=40$ is represented in terms of the mean cosine similarity (see Fig. \ref{fig:cosine similarity}), which is the metric used to compare the performance of the reconstructed velocity fields with respect to the reference ones. Three different energy reconstructions are evaluated. It must be remarked that the cosine similarity of the reconstruction with the \ac{GM} and the interpolator is assessed with respect to the LOR based on the same number of modes. In all cases, the \ac{GM} shows superior performance with respect to the interpolator, independently of the time distance from the initial condition. 

Regarding the effect of energy truncation on the reconstruction accuracy, different behaviours can be observed for the fluidic pinball and the jet flow. In the case of the fluidic pinball, the reconstruction with $70\%$ energy presents a lower performance in reconstruction with respect to higher energy, as 3 modes are not sufficient to define the full dynamics. Moreover, it can be noted that the reconstruction with $90\%$ provides a worse performance compared to $80\%$ due to the introduction of numerical errors from the integration of the lower-energy reconstructed mode. The reconstruction with $80\%$ of the energy provides the optimal reconstruction among the tested cases, as it contains sufficient number of modes to provide a reconstruction comparable to the LOR while not including modes degrading the performance of the model. In the case of the jet flow, it is observed that the cosine similarity is decreased when considering more modes in the reconstruction. Thus, the performance is slighted worse due to the introduction of larger numerical errors in the modes with less energy content, which are expected to be more affected by three-dimensional effects that cannot be captured by the \ac{GM} based on 2D data. 

The root mean square error of each mode for every case is reported in Fig. \ref{fig:error-modes} to analyse the reconstruction error of each of the modes in terms of the total energy of the truncated reconstruction. Thus, the error of lower-order modes decreases when the total reconstructed energy is increased. On the downside, the reconstruction error increases for higher order modes due to the loss of information of the interaction with smaller-scale structures, associated to truncated modes. This effect is more dominant for the fluidic pinball due to the small number of modes required to reconstruct the main dynamics of the flow. In the case of the jet flow, the wide energy distribution of the modes, reduces significantly the effect of truncation error in the higher energy modes. Furthermore, the effect is amplified with increasing snapshot separation due to the introduction of numerical errors from the integration. This effect is better observed in Fig. \ref{fig:FOM70}, which compares the reconstruction of $70\%$ of the total energy with the model truncated at $70\%$ and $90\%$ of the energy. The \ac{LOR} with $70\%$ energy is also given as reference. The main effect that can be observed, is the reduction in reconstruction error when considering more energy in the model. This reduction is more significant in the case of the fluidic pinball where the reconstruction error is minimised, while in the jet flow, it is slightly smaller. Furthermore, in terms of snapshot reconstruction, it can be noted that the model with $90\%$ energy, the structures are more similar to the reference in terms of shape and velocity intensity.

\section{\label{sec:conclusion}Conclusion}
A low-dimensional physics-informed model has been proposed for the temporal super-sampling of \ac{NTR} \ac{PIV} velocity fields, based on a Galerkin-POD projection of the Navier-Stokes equations. Temporal resolution enhancement is attained by inference of a flow dynamical model previously projected onto a low-order linear space. Integration of the dynamics is carried out through a forward-backwards weighting process, in between available snapshots. Projection of the predicted states onto the original high-dimensional space allows the retrieval of a more detailed and time-continuous representation of the flow field.

The method is evaluated on two datasets, the first one corresponding to a 2D simulation of a fluidic pinball, and the second one to the planar \ac{PIV} of a jet flow. The reconstruction with the \ac{GM} is then compared to a spline interpolator of the \ac{POD} temporal modes. Our results demonstrate that the \ac{GM} can effectively increase the temporal resolution over several flow characteristic times. When increasing the number of modes, the overall truncation error is reduced. Additionally, it is observed that adding more modes with small energy content degrades the reconstruction performance due to the introduction of errors coming from the least energetic modes. These findings are consistent with the increasing complexity reported in the literature for modelling multi-scale flows with high structure interactions.

The study demonstrates that the \ac{GM} is effective in reconstructing the dynamics of turbulent flows when no \ac{TR} measurements are available. It must be remarked, however, that the method is demonstrated here with two test cases with relatively small Reynolds number and compact \ac{POD} spectrum. This is expected to be an ideal condition for the effectiveness of methods based on Galerkin projection on \ac{POD} basis. We thus envision further work to explore the use of these methods at higher Reynolds numbers and in spectrally-richer flows. Further work may include a more extensive analysis on the reconstruction accuracy involving the pressure term for multi-frame PIV systems, as well as mitigation strategies towards error propagation during forward and backward numerical integration processes.

\section*{Acknowledgments}
This project has received funding from the European Research Council (ERC) under the European Union’s Horizon 2020 research and innovation program (grant agreement No 949085). Views and opinions expressed are however those of the authors only and do not necessarily reflect those of the European Union or the European Research Council. Neither the European Union nor the granting authority can be held responsible for them.

\bibliographystyle{RS}
\bibliography{references}  






\end{document}